\documentclass[10pt]{iopart}
\usepackage{iopams}
\usepackage{hyperref}
\usepackage{graphicx}
\usepackage{color}
\begin{document}

\title[Interacting Brownian Motion with resetting]{Interacting Brownian Motion with Resetting}

\author{R. Falcao$^1$ \footnote{on leave at: SUPA, School of Physics and Astronomy, University of Edinburgh, Mayfield Road,
Edinburgh EH9 3JZ, UK} and  M. R. Evans$^2$} 

\address{$^1$ Departamento de F\'\i sica e Matem\'atica, Universidade Federal de S\~ ao Jo\~ao del-Rei, 36420-000 Ouro Branco, MG, Brazil}
\address{$^2$ SUPA, School of Physics and Astronomy, University of Edinburgh, Mayfield Road,
Edinburgh EH9 3JZ, UK}
\ead{rfalcao@ufsj.edu.br,  m.evans@ed.ac.uk}

\begin{abstract}
We study two Brownian particles in dimension $d=1$, diffusing under an interacting resetting mechanism to a fixed position. The particles are subject to a constant drift, which biases the Brownian particles toward each other. We derive the steady-state distributions and study the late time relaxation behavior to the stationary state.
\end{abstract}


\vspace{2pc}

\submitto{\JSTAT}

\maketitle

\section{Introduction}

Resetting is the action of interrupting a continuously
evolving process and instantaneously bringing it back  to a predetermined state
to allow the process to restart.  When applied stochastically such an
action may represent a wide variety of phenomena. For example, in the
search for a lost object, after an unsuccessful period of search one
often spontaneously returns to the start position and restarts the
search from there.  Beyond the search strategies similar notions of
stochastic resetting are also found in:  population dynamics, where a
random catastrophic event can cause a drastic reduction in the
population, resetting it to some previuos value \cite{PRSM2010,
  MZ1999}; economics, where a financial crash may reset the price of
stocks assets to some predetermined value \cite{S2003};  biological
contexts, where organisms use stochastic resetting or switching
between different phenotypic states to adapt to fluctuating
environments \cite{PRSM2010, KL2005, KKBL2005}.

The Brownian particle  with stochastic resetting to an initial position  is an archetypal  realization of a resetting process\cite{MM2011}.
In the absence of a resetting mechanism the motion is purely diffusive and position of a Brownian particle has a Gaussian distribution with a variance that grows linearly in time, implying the absence of a steady state on an infinite system.  In the presence of resetting, the diffusive spread is opposed by the resetting leading to  confinement around the initial position  and  a nonequilibrium stationary state (NESS) is attained. 
In this NESS probability currents are non-zero and detailed balance does not hold---a non-vanishing steady-state current is directed towards the resetting position.

The study of NESS is of fundamental importance in statistical physics
\cite{KRB2010,DR2001}.  Generally, the existence of currents
allows a broader range of phenomena than in equilibrium. For example
boundary-induced phase transitions and generic long-range
correlations %(Include reference???).%
The resetting paradigm furnishes a simple way of
generating a non-equilibrium state by keeping the system away from any
equilibrium state by the constant reset to the initial condition.
Thus, the resetting paradigm provides a convenient framework to study
the properties of non-equilibrium states.  In more general
non-equilibrium contexts resetting has also been studied in
fluctuating interfaces \cite{MGS2014,GN2016}, and in a coagulation-diffusion
process \cite{DHP2015} and in reaction processes \cite{RSU2015}.  The
large deviations of time-additive functions of Markov processes with
resetting is considered in \cite{MST2015} and the thermodynamics of
resetting processes far from equilibrium in \cite{FGS2016}. A universal result for the fluctuations in first passage times of an optimally restarted process is obtained in \cite{R2016}. 

A number of generalisations of simple diffusion with resetting have been made: 
the  $d$-dimensional case has been considered in \cite{MM2014},
spatial resetting distributions and spatially-dependent resetting rates are  studied in \cite{MM2011a}.
The properties of the non-equilibrium steady state have been studied in the presence of a potential \cite{P2015} and in a bounded domain \cite{CS2015}.  The dynamics of reaching the NESS was studied and a dynamical phase transition was found in \cite{MSS2015b}. In the context of random walks,
resetting in continuous-time random walks \cite{MV2013, MC2016}, in L\'evy flights \cite{KMSS2014}, in random walks with exponentially distributed flights of constant speed \cite{CM2015} have been considered

In all of these  works the resetting mechanism occurs through  an {\em external} mechanism,
modelled in the continous time case  as a Poisson process. That is, resetting occurs when an external bell rings generating an exponential distribution of waiting times between resets.
Some generalisations to non-Markovian processes where the waiting time between resets
is non-exponential have been considered \cite{EM2016,PKM2016,NG2016}, a deterministic
resetting for multiple searchers is studied in \cite{BBR2016}
and a generalisation to where the reset position depends on the internal dyanamics
such as to the current maximum of a random walk \cite{MSS2015a} 
or to a position selected from a resetting distribution \cite{MM2011a}
have been studied.
However the source of resetting has always remained external.

In this paper we seek to extend the field of study by introducing
resetting that is triggered by the {\em internal} dynamics of a system.
Instead of using a constant rate, or using any
predetermined  waiting time distributions between  resetting events,
the resetting is triggered  through interactions between the constituent particles.

In order to study such an interaction-driven resetting mechanism we
propose in this work a toy model that consists of two Brownian
particles in one dimension subject to mutual attraction and resetting
to the initial position every time they are about to collide.  Thus
the two particle system contracts stochastically then sudden dilation
to the intial configuration occurs when the particles are adjacent.
This resetting mechanism is essentially different from all the
previous studies.  Our toy model is simple enough to allow an exact
solution yet rich enough to  yield a non-trivial NESS.  In particular, our solution
allows study of the effective resetting rate induced by the
interactions.

The paper is organised as follows. In Section 2 we define our toy
model of two random walkers with a hard-core interaction that triggers
resetting.  In Section 3 we present a solution of the master equation
for the time-dependent probability distribution using a
self-consistent initial value Green function technique.  In section 4
we determine the stationary state. In section 5 we go on to compute
the time-dependent probability distribution and induced resetting
rate, presenting approximations accurate in different regimes. We
conclude in section 6.

\section{The Model}

We start with a lattice model to make clear the resetting mechanism, then we take the continuous limit and study the model in this limit. The lattice model consists  of two asymmetric random walkers moving on a one-dimensional lattice, the left walker has a higher probability to jump to the right and the right walker a higher probability to jump to the left, the walkers are also subject to a resetting mechanism which relocates both walkers in their initial positions when they are about to collide.

Let $x_L(t)$ and $x_R(t)$ denote the position of the left/right walker at step $t$, $x_L(0)=-l$ and $x_R(0)=l$, $l\in \mathbb{N}$ are the initial positions of each walker.    

The positions $x_L(t)$ and $x_R(t)$ evolves with time via the following stochastic dynamics: At any given time step $t$, if the position
$x_R(t)- x_L(t)>2$, then in the next time step the right walker moves to the left with probability $1/2+\epsilon$ and to the right with probability $1/2-\epsilon$ and the left walker moves to the right with probability $1/2+\epsilon$ and to the left with probability $1/2-\epsilon$. If the position $x_R(t)-x_L(t)=2$  then in the next step the walkers moves right or left or both reset to their initial positions. This dynamics can be interpreted as two attracting random walkers with resetting, and is defined by the following evolution rules:   

\begin{figure}
\begin{center}
\begin{tabular}{cc}
(a)& (b) \\
&\\
\includegraphics[width=6.1cm]{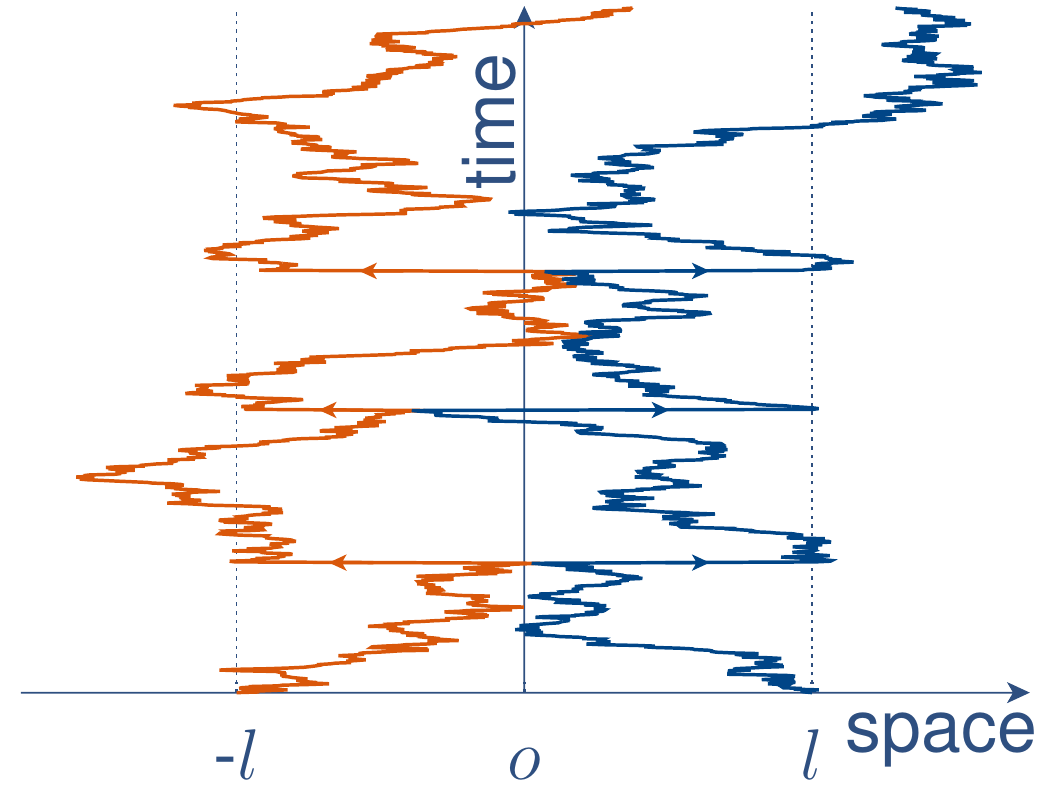}  &\includegraphics[width=6.1cm]{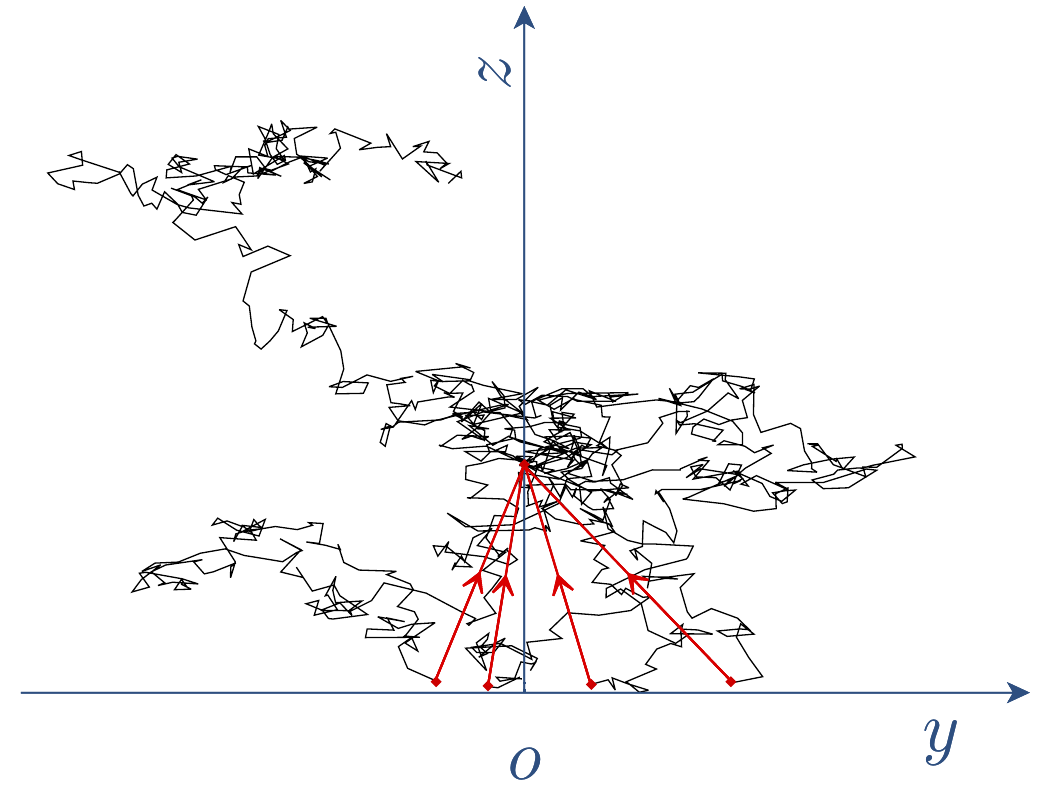} 
\end{tabular}
\caption{(a) Space-time trajectory for two one-dimensional Brownian particles with interaction triggering resetting mechanism (b) trajectory in the $yz$-plane in the new coordinates system equations (\ref{defyz}).}
\end{center}
\end{figure} 

if $x_L-x_R>2$
\begin{eqnarray}
(x_L,x_R) \rightarrow \left \{  \begin{array}{l} (x_L\!+\!1,x_R\!+\!1), ~~ \textrm{with probability} ~ 1/4-\epsilon^2, \\
                                 (x_L\!-\!1,x_R\!-\!1), ~~\textrm{with probability} ~ 1/4-\epsilon^2,  \\
                                 (x_L\!+\!1,x_R\!-\!1), ~~\textrm{with probability} ~ (1/2+\epsilon)^2, \\
                                 (x_L\!-\!1,x_R\!+\!1), ~~\textrm{with probability} ~ (1/2-\epsilon)^2, \\
                                \end{array} \right .\label{dynamicsx1}
\end{eqnarray}
and if $x_L-x_R=2$
\begin{eqnarray}
(x_L,x_R) \rightarrow \left \{  \begin{array}{l} (x_L\!+\!1,x_R\!+\!1), ~~ \textrm{with probability} ~ 1/4-\epsilon^2, \\
                                 (x_L\!-\!1,x_R\!-\!1), ~~\textrm{with probability} ~ 1/4-\epsilon^2, \\
                                 (-l,l), ~~\textrm{with probability} ~ (1/2+\epsilon)^2, \\
                                 (x_L\!-\!1,x_R\!+\!1), ~~\textrm{with probability} ~ (1/2-\epsilon)^2. \\
                                \end{array} \right .\label{dynamicsx2}
\end{eqnarray}
Since we have $x_R>x_L$, it is convenient to define two new variables:
\begin{equation}
y =(x_R + x_L)/2, \qquad z =(x_R - x_L)/2 \label{defyz}.
\end{equation}
Here we may think of $y$ as the centre-of-mass of the pair of particles and 
$z$ represents (half) the separation.
In terms of $y$ and $z$, the dynamics in equations  (\ref{dynamicsx1}, \ref{dynamicsx2}) are translated into: 

if $z>1$
\begin{eqnarray}
(y,z) \rightarrow \left \{  \begin{array}{l} (y,z\!+\!1), ~~ \textrm{with probability} ~ (1/2-\epsilon)^2, \\
                                 (y,z\!-\!1), ~~\textrm{with probability} ~ (1/2+\epsilon)^2, \\
                                 (y\!-\!1,z), ~~\textrm{with probability} ~ 1/4-\epsilon^2, \\
                                 (y\!+\!1,z), ~~\textrm{with probability} ~ 1/4-\epsilon^2. \\
                                \end{array} \right .\label{dynamicsz1}
\end{eqnarray}

and if $z=1$
\begin{eqnarray}
(y,z) \rightarrow \left \{  \begin{array}{l} (y,z\!+\!1), ~~ \textrm{with probability} ~ (1/2-\epsilon)^2, \\
                                 (0,l), ~~\textrm{with probability} ~ (1/2+\epsilon)^2, \\
                                 (y\!-\!1,z), ~~\textrm{with probability} ~ 1/4-\epsilon^2, \\
                                 (y\!+\!1,z), ~~\textrm{with probability} ~ 1/4-\epsilon^2. \\
                                \end{array} \right .\label{dynamicsz2}
\end{eqnarray}

Let $P(y,z,t)$ denote the joint probability distribution of $(y,z)$ at the $t$-th time step. Evidently, $P((x_R+x_L)/2,(x_R-x_L)/2,t)=Q(x_L,x_R,t)$ where $Q(x_L,x_R,t)$ is the joint probability distribution of
the positions  $(x_L,x_R)$. Using the dynamics in equations (\ref{dynamicsz1}) and (\ref{dynamicsz2}), it is easy to write down the master equation for $P(y,z,t)$ as
\begin{eqnarray}\fl
P(y,z,t)=\left(1/2+\epsilon\right)^2P(y,z+1,t-1)+\left(1/2-\epsilon\right)^2P(y,z-1,t-1) \nonumber \\ 
+\left(1/4-\epsilon^2\right)\left(P(y+1,z,t-1)+P(y-1,z,t-1)\right )\nonumber \\
+\left(1/2+\epsilon\right)^2\delta_{y,0}\delta_{z,l}\sum_{k}P(k,1,t-1)\;.
\end{eqnarray}

In the continuous time and space limit, i.e. changing the time step to $\Delta t$ and the lattice size to $\Delta x$,  we obtain the following Master equation
\begin{equation}
D \nabla^2 P(y,z,t)+v\frac{\partial P(y,z,t)}{\partial z}+f(t)\delta(y)\delta(z-l)=\frac{\partial P(y,z,t)}{\partial t} \label{diff}
\end{equation}
where $\nabla^2=\frac{\partial^2}{\partial y^2}+\frac{\partial^2}{\partial z^2}$, $v=\lim_{\Delta x, \Delta t, \epsilon\rightarrow 0}\frac{2\epsilon\Delta x}{\Delta t}$, $D=\lim_{\Delta x, \Delta t \rightarrow 0}\frac{(\Delta x)^2}{4\Delta t}$ and
\begin{eqnarray}
f(t)=D\int_{-\infty}^\infty\frac{\partial P(y',0,t)}{\partial z}\rmd y'\label{fdef}.
\end{eqnarray}
In  (\ref{diff}) the terms on the first term on left hand side represents the diffusive
behaviour of both centre-of-mass $y$ and half-separation $z$; the second term  represents
the drift in $z$ due to particle moving toward each other with a bias $v$ and the third term  represents  resetting to  $z=l$ , $y=0$ with rate 
$f(t)$. 
Thus $f(t)$ is the effective resetting which is due to the diffusive probability current
into the line $z=0$.
The rate is 
determined by the integral of the derivative of $P$ at  $z=0$
(\ref{fdef}).
Assuming $\epsilon>0$ implies that the bias  $v>0$; if $v<0$ there is no stationary state for this problem, since the two walkers tends to walk apart. With the change of variables to $(y,z)$  the  two attracting Brownian particle with resetting problem reduces to  that of  a $2$-$d$ Brownian particle with an absorbing boundary and a time-dependent source term.

\section{Green Function Solution}
To obtain the solution to (\ref{diff}, \ref{fdef}) we use a Green function approach.
The first step is to find the initial value Green function for the homogeneous problem 
\begin{equation}
D\nabla^2 G(y,z,t)+v\frac{\partial G(y,z,t)}{\partial z}=\frac{\partial G(y,z,t)}{\partial t} 
\label{Green}
\end{equation}
subject to the initial condition and boundary conditions:
\begin{equation}
G(y,z,0)=\delta(y)\delta(z-l), \qquad G(y,0,t)=0. \label{Gbc}
\end{equation}
Then the full time-dependent solution of (\ref{diff}) can be written down as
\begin{equation}
P(y,z,t)=G(y,z,t)+\int_{0}^{t}G(y,z,t-t')f(t')\rmd t' \label{fulltime}
\end{equation}
where the resetting rate  $f(t)$ given by (\ref{fdef}) is obtained self-consistently as we shall detail below.
The first term in (\ref{fulltime}) is the contribution from trajectories where no resetting has occurred
and the second term is contributions where the last reset occurred at time $t'$.

To obtain the Green function we
take the Laplace-Fourier transform of (\ref{Green})  with respect to $t$ and $y$,
\begin{equation}
\tilde{G}(y,z,s)=\frac{1}{\sqrt{2\pi}} \int_{-\infty}^{+\infty}\rme^{iky} \int_{0}^\infty \rme^{-st}G(y,z,t)\rmd t\rmd y
\end{equation}
to obtain the following equation
\begin{equation}
D\frac{d^2 \tilde{G}(k,z,s)}{dz^2}+v\frac{d\tilde{G}(k,z,s)}{dz}-(Dk^2+s)\tilde{G}(k,z,s)=-\tilde{G}(k,z,0)\label{diffeq}
\end{equation}
where $\tilde{G}(k,z,0)=\frac{\delta(z-l)}{\sqrt{2\pi}}$. The homogeneous solution of equation (\ref{diffeq}) is
\begin{equation}
\tilde{G}_h=A\rme^{\lambda^{+}z}+B\rme^{\lambda^{-}z}
\end{equation}
where
\begin{equation}
\lambda^{\pm}=-\frac{v}{2D}\pm\left (\frac{v^2}{4D^2}+\frac{Dk^2+s}{D}\right )^{1/2}
\end{equation}
and a particular solution of the inhomogeneous equation  (\ref{diffeq}) is
\begin{equation}
\tilde{G}_i=\frac{H(l-z)}{2\Lambda D\sqrt{2\pi}}\left(\rme^{-\lambda^{+}(l-z)}-\rme^{-\lambda^{-}(l-z)}\right )
\end{equation}
where $H(z)$ is the Heaviside function 
\begin{eqnarray}
 H(z)=\cases{0, \qquad z<0 \\
 \frac{1}{2}, \qquad z=0 \\
 1, \qquad z>0}
\end{eqnarray}
and $\Lambda=\sqrt{\frac{v^2}{4D^2}+\frac{Dk^2+s}{D}}$. Since $\lambda^{+}>0$ the only way to ensure that $\tilde{G}$ remains finite when $z\rightarrow \infty$ is to put $A=0$ , and to obey the boundary condition $\tilde{G}(k,0,s)=0$ we obtain 
\begin{equation}
B=\exp\left(\frac{vl}{2D}\right)\frac{\sinh(\Lambda l)}{\sqrt{2\pi}\Lambda D}
\end{equation}
and the solution is given by
\begin{eqnarray}
\tilde{G}(k,z,s)=\cases{  \rme^{\frac{v}{2D}(l-z)} \left ( \frac{\rme^{-\Lambda(l-z)}-\rme^{-\Lambda(l+z)}}{\sqrt{8\pi}\Lambda D}\right ),\qquad z<l\\
                            \rme^{\frac{v}{2D}(l-z) } \left ( \frac{\rme^{-\Lambda(z-l)}-\rme^{-\Lambda(l+z)}}{\sqrt{8\pi}\Lambda D}\right ),\qquad z\geq l.\\
                          }
\end{eqnarray}
Inverting the Laplace transform we obtain
\begin{equation}
\mathcal{L}^{-1}\left \{\tilde{G}(k,z,s)\right \}=\frac{\rme^{\frac{v}{2D}(l-z)}}{\sqrt{8 \pi^2 D t}}\left(\rme^{-\frac{(z-l)^2}{4Dt}-ct}-\rme^{-\frac{(z+l)^2}{4Dt}-ct} \right)
\end{equation}
where $c={v^2}/{4D}+Dk^2$. Finally, inverting the Fourier transform we obtain
the solution of (\ref{Green},\ref{Gbc})
\begin{eqnarray}\fl
G(y,z,t)=\frac{1}{4\pi Dt}\left[\exp\left (-\frac{\left(z-l+vt\right)^2+y^2}{4Dt}\right ) \right .\nonumber \\ 
\left . -\exp\left (-\frac{\left(z+l+vt\right)^2+y^2}{4Dt}\right )\exp\left(\frac{vl}{D}\right)\right]\;. \label{hsolution}
\end{eqnarray}
One can understand the second term in (\ref{hsolution}) as  representing the ``image'' contribution to the solution,  the drift velocity of the image need to be in the same direction as that of the original particle, and the additional exponential factor in the image term is to ensure that the boundary condition $G(0,y,t)=0$ is obeyed. 

\section{Stationary State}

Due to the drift  of the particles towards each other (drift in the $z$ direction) and
the resetting mechanism (the jump from the boundary $z=0$ to the position $(0,l)$) a stationary state
of (\ref{diff})  is expected to exist for this system.
Now $G(y,z,t) \to 0$ as $t \to \infty$ so the stationary state 
of (\ref{fulltime}) is given by
\begin{equation}
P_{st}(y,z)= \lim_{t\rightarrow\infty} \int_{0}^{t}G(y,z,t-t')f(t')\rmd t' \label{stst}\;.
\end{equation}
 In the stationary state the function $f(t)$, given by (\ref{fdef}), must be independent of $t$ so to obtain the stationary state we need only  solve the problem for a constant $N$ in place of the function $f(t)$, where  this constant should be $N=\lim_{t \rightarrow \infty} f(t)$. In this way the stationary solution is given by 
\begin{equation}\fl
P_{st}(y,z)=\lim_{t\rightarrow\infty}\frac{N}{4\pi D}\int_{0}^t \tau^{-1}\rme^{ \frac{v}{2D}(l-z)-\frac{v^2\tau}{4D}}\left(\rme^{-\frac{\left(z-l\right)^2+y^2}{4D\tau}}-\rme^{-\frac{\left(z+l \right)^2+y^2}{4D\tau}}\right)\rmd \tau \;.
\end{equation}
Using the following change of variables
\begin{equation}
a_{\mp}=\frac{(z\pm l)^2}{4D}+\frac{y^2}{4D};~~~~b=\frac{v^2}{4D},~~~u=\frac{a_{\pm}}{\tau}
\end{equation}
where $a_{+}$ is used in the first integral and $a_{-}$ in the second, we obtain
\begin{eqnarray}\fl
P_{st}(y,z)=\lim_{t\rightarrow\infty}\frac{N}{4\pi D}\rme^{-\frac{v(z-l)}{2D}}\left [ \int\limits_{0}^{\infty}\exp\left(-u-\frac{ba_{+}}{u}\right)\frac{\rmd u}{u}-\int\limits_{0}^{\infty}\exp\left(-u-\frac{ba_{-}}{u}\right)\frac{\rmd u}{u}\right . \nonumber \\
\left.-\int\limits_{0}^{a_+/t}\exp\left (-u-\frac{ba_{+}}{u}\right )\frac{\rmd u}{u} +\int\limits_{0}^{a_{-}/t}\exp\left (-u-\frac{ba_{-}}{u}\right )\frac{\rmd u}{u}\right ].
\end{eqnarray}
Taking the limit $t\to \infty$ we obtain
\begin{equation}\fl
P_{st}=\frac{N}{2\pi D}\rme^{-v(z-l)/2D}\left[K_0\left( \frac{v}{2D}\sqrt{(z-l)^2+y^2} \right ) - K_0\left( \frac{v}{2D}\sqrt{(z+l)^2+y^2} \right ) \right] \label{sst}
\end{equation}
where $K_0$ is the zero-order modified Bessel function of the second kind  and $N$ is
to be  determined self-consistently. This can  be done through the
conservation of probability $\int P_{st}(y,z) \rmd y \rmd z =1$
and one obtains $N= v/l$.

In what follows it will be useful to define the P\'eclet number 
\begin{equation}
P_e=vl/2D
\label{Pe}
\end{equation}
which characterizes the relative importance between diffusion and convection in a biased diffusive process. 

\begin{figure}[!ht]
\begin{center}
\begin{tabular}{cc}
(a) &(b) \\
\includegraphics[height=5.73cm]{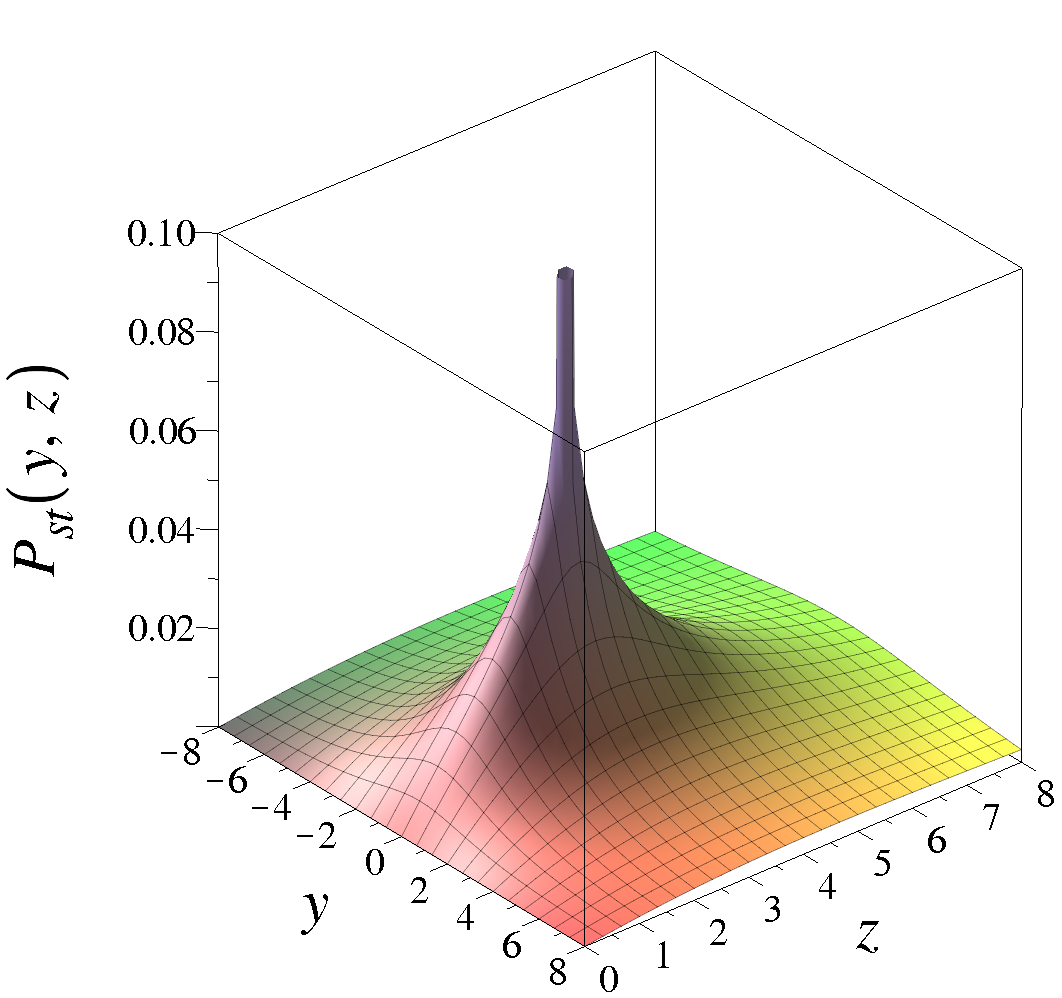} & \includegraphics[height=5.73cm]{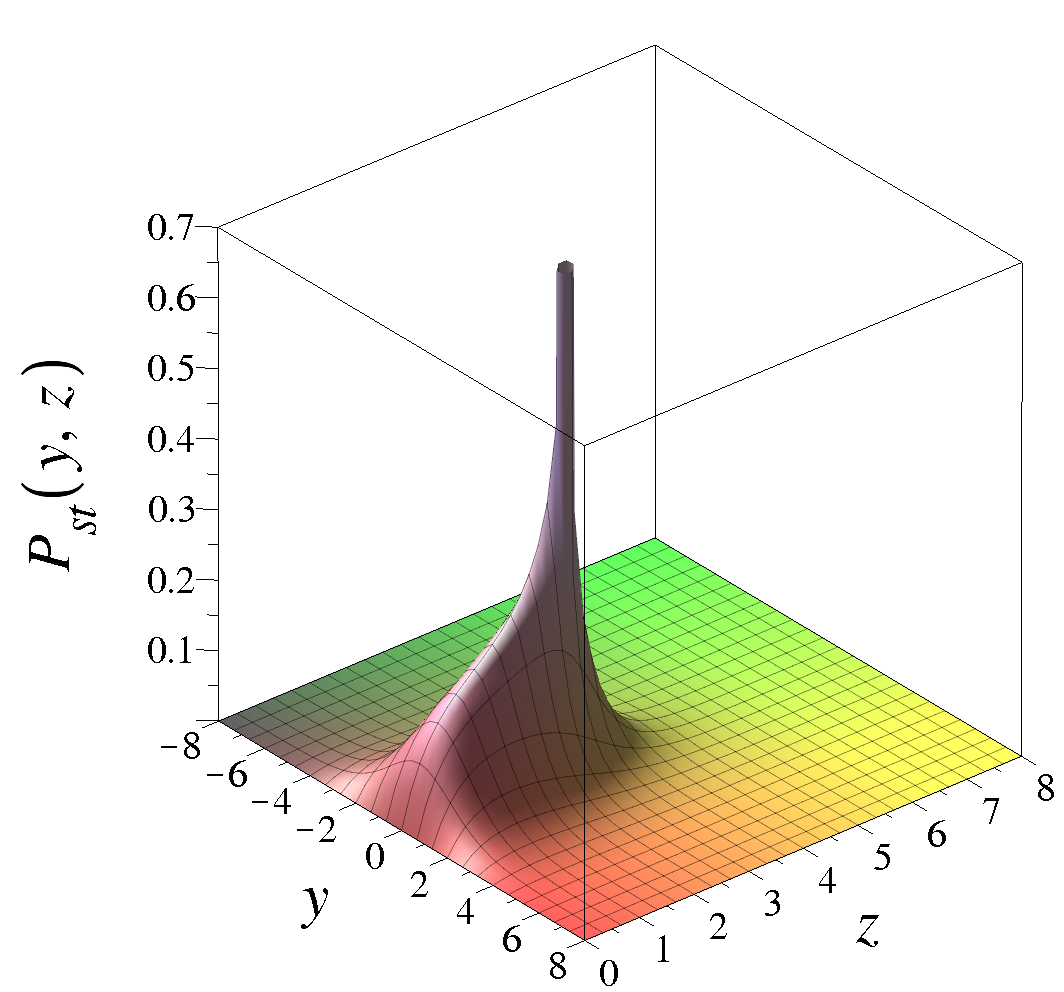}\\
& \\
(c) &(d) \\
\includegraphics[width=6.1cm]{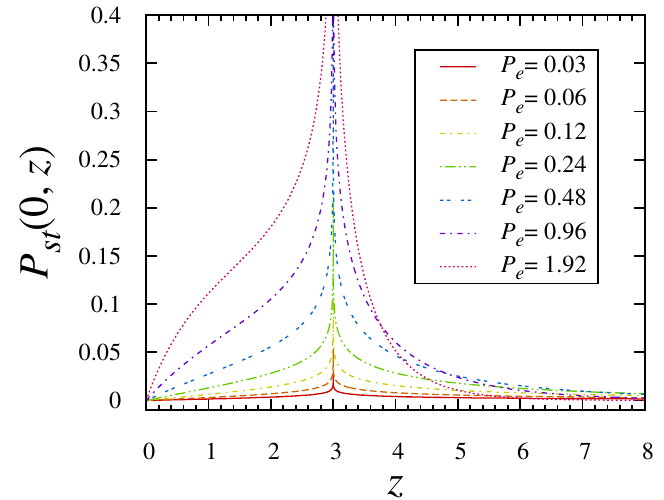} & \includegraphics[width=6.1cm]{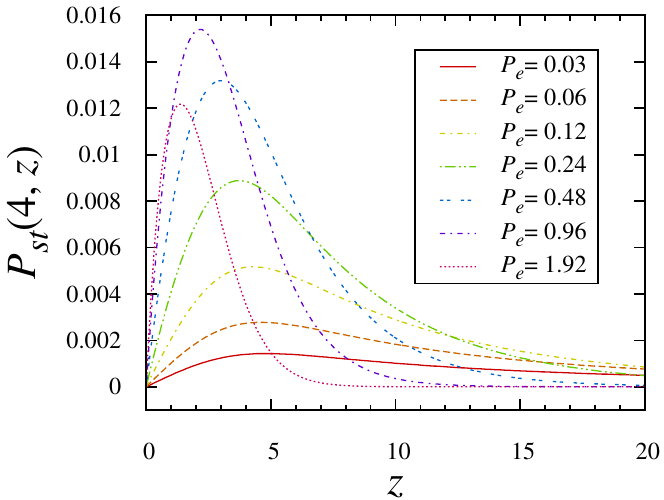}
\end{tabular}
\caption{Stationary probability density for different P\'eclet numbers, (a) $P_e=0.3$, (b) $P_e=3$. Cross section of the probability density (c) for $y=0$ , (d) for $y=4$ for different P\'eclet numbers. \label{fig1}}
\end{center}
\end{figure}

Figure \ref{fig1} shows the stationary probability for two different
values of the  P\'eclet number. We observe in Figure \ref{fig1}-b that for large Peclet number the steady state is quite asymmetric with respect to $z$ (the half-separation) which reflects the fact that the drift is strong and after resetting to the  initial value of $z$, $z$ tends strongly to decrease.
Also due to the absorbing  boundary condition at $z=0$ there is a sudden variation of the probability density next to the $y$ axis, creating a  kind of boundary layer.  On the other hand, for small Peclet number where the drift is weak the stationary state is almost symmetric around the resetting point. We can see in Figure \ref{fig1}-d that for small Peclet number the maximum of the probability density, for example in the cross section $y=4$, can occur at values of $z$ greater than the resetting point which is at  $z=3$  in this case.   

We also note from formula (\ref{sst}) and Figure \ref{fig1} that there is a
logarithmic  singularity in the distribution at $z=l$, $y=0$ (the resetting
point). Explicitly, if we set  $y=0$ and $z= l +\epsilon$
then for $|\epsilon| \ll 1$
\begin{eqnarray}
P_{st}(0,l+\epsilon) \sim
\frac{N}{2\pi D}\ln \left(\frac{4D}{v|\epsilon|}\right)
\label{stsin}
\end{eqnarray}
where we have used 
\begin{equation}
K_0\left(x\right)\sim -\ln \left(\frac{x}{2}\right)-\gamma,\qquad \textrm{for } 0<x\ll1.\label{bessels}
\end{equation}

It is interesting to compare the singularity  in (\ref{stsin}) with results for
simple diffusion with resetting.
As we have seen the present model reduces to  
a two dimensional walker (in the $y$--$z$ plane) with resetting when the walker reaches the line $z=0$.
The simple diffusion with  stochastic resetting at constant rate has a curious behaviour: the  stationary probability density has a cusp at the resetting point for all dimension $d$ except for $d=2$,  where the probability density diverges logarithmic at the resetting point \cite{MM2014}.
Therefore the logarithmic singularity that we obtain is consistent 
with this two-dimensional behaviour.

We can also see what happens in a  problem similar to ours but in one dimension, i.e. a Brownian particle with a drift towards the origin and this particle is resetting to the initial position when approaching the origin. The master equation for this problem is given by   
\begin{equation}
D\frac{\partial^2P(x,t)}{\partial x^2}+D\delta(x-l)\frac{\partial P(0,t)}{\partial x}+v\frac{\partial P(x,t)}{\partial x}=\frac{\partial P(x,t)}{\partial t} \label{diff1d}
\end{equation}
and the stationary state solution of equation (\ref{diff1d}) is a simple exercise,  and is given by 
\begin{eqnarray}
P_{st}(x)=\cases { \begin{array}{l} 
                  \frac{1}{l}\left(1-\rme^{-\frac{vx}{D}}\right ), \qquad 0<x\leq l \\ 
                  \frac{1}{l}\rme^{-\frac{vx}{D}}(\rme^{\frac{vl}{D}}-1)\qquad x>l \\
                 \end{array} }
\end{eqnarray}
We observe the existence of a {\em cusp} in $x=l$ instead of a logarithmic divergence  observed in equation (\ref{sst}) when $(y,z)\rightarrow (0,l)$ so the change from a cusp in one dimension to  a logarithmic divergence in two dimensions is also observed in our model.

\subsection{Small v and Scaling Limit of Stationary State}
It is of interest to explore the limiting cases where  diffusion dominates
the resetting process. We shall consider two limits: small bias $v\to0$ and a scaling limit
$v\to 0$, $ z\to \infty$ with $vz$ constant.

\begin{figure}[!ht]
\begin{center}
 \includegraphics[width=9cm]{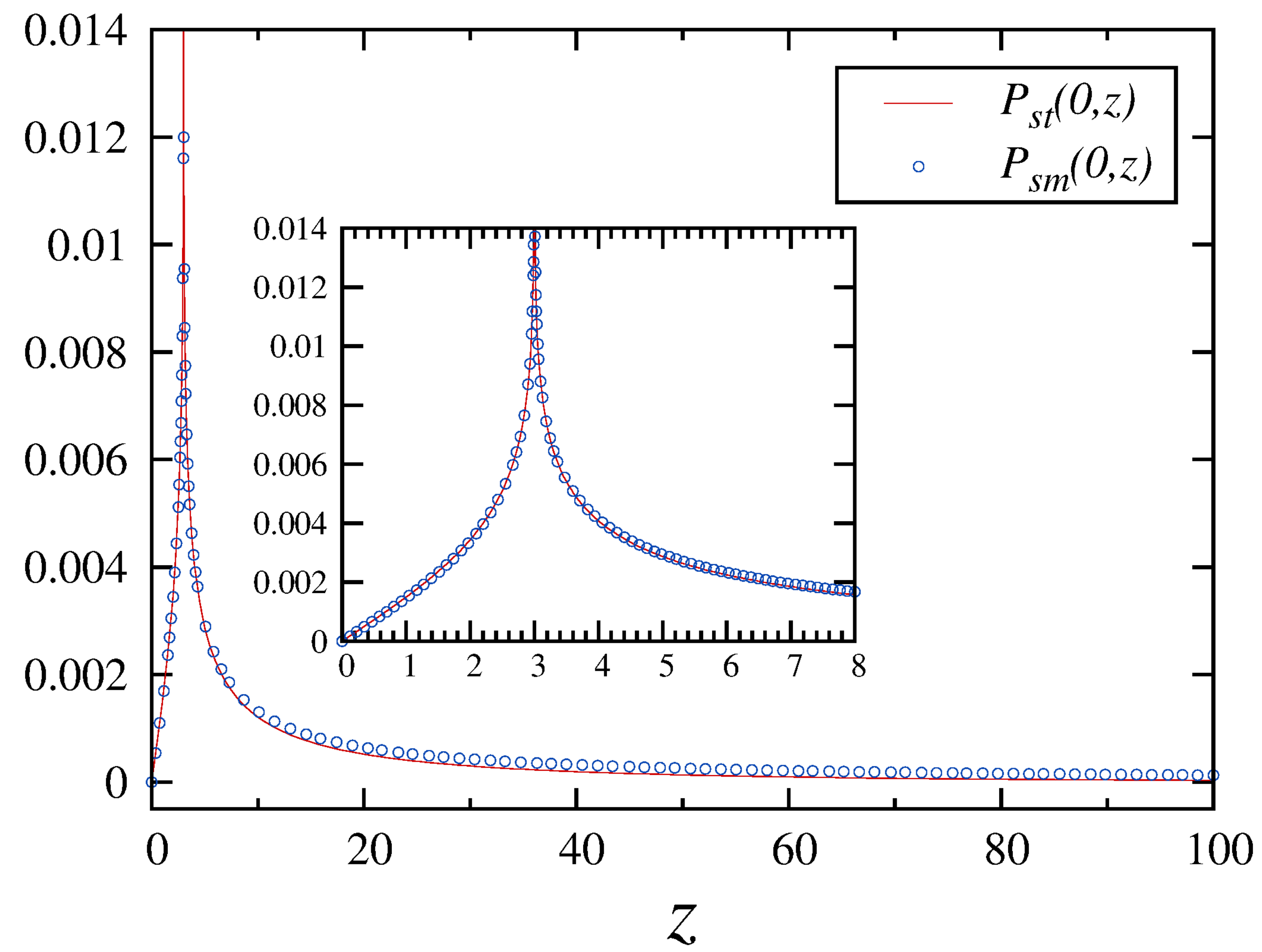} 
\caption{ The line is the stationary solution $P_{st}$ for $y=0$ and the dashed line is the approximation given by Eq. (\ref{small})}\label{Figsmall}
\end{center}
\end{figure}

The small $v$ regime of the stationary state is obtained using the small argument approximation for the modified Bessel function (\ref{bessels}).
Keeping $\sqrt{(z\mp l)^2+y^2}$ finite and making $v\rightarrow 0$ in equation (\ref{sst}) we obtain using the approximation (\ref{bessels}),  
\begin{equation}
P_{st}(y,z)=\frac{v}{2\pi l D}\ln\left(\frac{\sqrt{(z+l)^2+y^2}}{\sqrt{(z-l)^2+y^2}}\right)
\end{equation}
in particular for $y=0$ we have
\begin{equation}
P_{st}(0,z)=\frac{v}{2\pi l D}\ln\left(\frac{z+l}{|z-l|}\right) \label{small}
\end{equation}
in the inset of Figure \ref{Figsmall} is possible to see that this approximation is accurate near the resetting point.

Looking in the scale where $vz=2D\xi$, in the limit of small drift, $v\rightarrow 0$, and large distances, $z\rightarrow \infty$ but keeping the product $vz=2D\xi$ constant. The stationary state for $y=0$ given by equation (\ref{sst}) can be rewritten as
\begin{equation}
P_{st}(0,z)=\frac{v}{2\pi l D}\textrm{e}^{-\xi+P_e}\left [K_0 \left(\xi-P_e\right)- K_0 \left(\xi+P_e\right) \right]
\end{equation}
using the fact that $K_0(\xi\pm P_e)\sim K_0(\xi)\mp K_1(\xi)P_e$, for $P_e\rightarrow 0$ we obtain in this scaling regime,
\begin{equation}
P_{st}(0,z)\sim\frac{2\textrm{e}^{\xi}\xi^2K_1(\xi)}{z^2}\;.
\end{equation}
Thus we see that  $z^2 P_{st}(0,z)$ is given by a non-trivial scaling function.

\section{Relaxation to the Stationary State}
We now turn to the full time-dependent solution equation of the Master equation (\ref{fulltime}).  The task is determine $f(t)$ in
(\ref{fulltime})  self-consistently.
Substituting (\ref{fulltime}) in  equation (\ref{fdef}) we obtain
\begin{equation}
f(t)=k(t)+\int k(t-t')f(t')dt' \label{selff}
\end{equation}
where 
\begin{equation}
k(t)=\int_{-\infty}^{\infty} \left .D\frac{\partial G(y',z,t)}{\partial z}\right |_{z=0}dy'=\frac{l}{\sqrt{4D\pi t^3}}\exp\left(-\frac{(tv-l)^2}{4Dt} \right)
\end{equation}
(we note in passing that this expression is equivalent to  the distribution of
first passage time   to the origin for a biased diffusion in one dimension starting at $l$ \cite{R2001}). Taking the Laplace transform of equation (\ref{selff}) we obtain
\begin{equation}
F(s)=\frac{K(s)}{1-K(s)}
\end{equation}
where $K(s)$ is the Laplace transform of $k(t)$ given by
\begin{equation}
K(s)=\exp\left ( \frac{l\left ( v-\sqrt{4Ds+v^2}\right)}{2D} \right )
\end{equation}
since $K(s)<1$, for $s>0$, we can use the sum of the geometric series to obtain 
\begin{equation}
F(s)=\sum_{n=1}^\infty \exp\left (\frac{nl\left ( v-\sqrt{4Ds+v^2}\right)}{2D} \right )\;.
\end{equation}
Inverting the Laplace transform term by term we find
\begin{equation}
f(t)=\sum_{n=1}^{\infty}\frac{nl}{\sqrt{4\pi D t^3}}\exp\left(-\frac{(tv-nl)^2}{4Dt}\right )=
\sum_{n=1}^{\infty}f_n(t)\;.
\label{fntdef}
\end{equation}
Thus we obtain  the full time-dependent solution of (\ref{diff})  as
\begin{equation}
P(y,z,t)=  G(y,z,t)+\int_{0}^{t} G(y,z,t-t')
\sum_{n=1}^{\infty}f_n(t')\rmd t' 
\label{fsol}
\end{equation}
where $G(y,z,t)$ is given by (\ref{Green})
and $f_n(t)$ is given by (\ref{fntdef}).

\subsection{Analysis of the resetting rate}

We have determined the  resetting rate $f(t)$ to be sum of contributions (\ref{fntdef}).
Each term of this sum is equal to first passage time  probability to the origin for a biased diffusion in one dimension starting at $nl$. 
Thus each term  $n$ in the sum corresponds to  reaching
$z=0$ at time  $t$ after $n-1$ previous resets.

In the extreme case  $D\to 0$ and $v$ finite, which is the deterministic limit,
the function $f$ is just a sum of delta functions spaced by a time interval $l/v$ which is the time necessary for the particle to reach the origin from the  initial (and relocation) position $l$. 
On the other hand in the limit $v\to 0$ we expect purely diffusive behaviour.
The function $f(t)$ interpolates between these two limits.

In the following we shall obtain  approximations to the resetting rate $f(t)$
in the regimes of small and large P\'eclet number given by (\ref{Pe}).

\subsubsection{Large $P_e$:} 
For large P\'eclet number (small $D$/large $v$) an approximation for the sum 
(\ref{fntdef}) is obtained using the identity
\begin{equation}
\sqrt{s}\, \vartheta_3(\pi a,e^{-\pi s})= \sum_{n=-\infty}^{\infty}\exp\left(-\frac{\pi(n-a)^2}{s}\right )
\end{equation}
where $\vartheta_3$ is a Jacobi Theta function \cite{PBM1998}.
Differentiating this formula with respect to $a$ we obtain
\begin{eqnarray}\fl
\sum_{n=-\infty}^{\infty}\frac{nl}{\sqrt{4 D\pi t^3}}\exp\left(-\frac{(tv-nl)^2}{4Dt}\right )=\frac{v}{l}\vartheta_3 \left(\frac{\pi vt}{l}, \exp\left(\frac{-4D\pi^2t}{l^2}\right)\right)\nonumber \\
+\frac{2D\pi}{l^2}\vartheta'_3 \left(\frac{\pi vt}{l}, \exp\left(\frac{-4D\pi^2t}{l^2}\right)\right)\;.
\end{eqnarray}
For $t>0$ and large $P_e$ the contribution of the negative values of $n$ in this sum is small 
so that $ \sum_{1}^{\infty} f_n(t)\simeq \sum_{-\infty}^{\infty} f_n(t)$ and we obtain the following approximation for $f(t)$
\begin{equation}\fl
f(t)\simeq \frac{v}{l}\vartheta_3 \left(\frac{\pi vt}{l}, \exp\left(\frac{-4D\pi^2t}{l^2}\right)\right)+\frac{2D\pi}{l^2}\vartheta'_3 \left(\frac{\pi vt}{l}, \exp\left(\frac{-4D\pi^2t}{l^2}\right)\right)\;.
\end{equation}
We now use identity \cite{PBM1998}
\begin{equation}
\vartheta_3(z,q)=1+2\sum_{n=1}^{\infty}q^{n^2}\cos(2nz)
\end{equation}
since $q= \exp( -4\pi^2 D t/l^2) \ll 1$ for large $Dt$ we can keep only the term  $n=1$  in the sum to obtain the large time behaviour. After some simple trigonometric indentities we obtain
\begin{equation}
f(t)\simeq f_l(t)= \frac{v}{l}\left\{ 1+2R\exp\left(-\frac{4\pi^2tD}{l^2}\right )\cos\left( \frac{2\pi vt}{l}+\phi \right)\right \} \label{largevf}
\end{equation}
where $R=\sqrt{1+(4D\pi/vl)^2}$ and $\phi=\arctan(4D\pi/lv)$. 
Thus, in the large Peclet Number regime and for large $Dt$ we have arrived at an approximation for the  effective resetting rate  that is a simple function of time: a damped harmonic oscillation deaying to constant rate $v/l$.

\subsubsection{Small $P_e$:} 
On the other hand, in the small P\'eclet number regime  the approximation $\sum_{1}^{\infty} f_n(t)\simeq \sum_{-\infty}^{\infty} f_n(t)$ does not hold, but 
instead a good approximation can be obtained by replacing the
sum  $\sum_{n=1}^{\infty}f_n(t)$  by an integral $\int_{0}^\infty f_n(t) dn$:
\begin{equation}\fl
f(t)\simeq f_s(t)=\int_{0}^\infty f_n(t) dn=\frac{v}{2l}+\frac{v}{2l}\textrm{erf}\left(\frac{v}{2}\frac{\sqrt{t}}{\sqrt{D}}\right)+\frac{\sqrt{D}}{l\sqrt{\pi t}}\textrm{e}^{-\frac{tv^2}{4D}}\;. \label{smallvf}
\end{equation}
In this case $f(t)$ has no oscillatory behaviour and $f_s(t)$ approaches  the stationary value $v/l$ monotonically from above.
It is interesting to note that
the decay rate depends on $v$, in contrast to the large P\'eclet number limit where the decay of the amplitude of the oscillation in (\ref{largevf}) does not depend on $v$. \\[1ex]

\begin{figure}[!ht]
\begin{center}
\begin{tabular}{cc}
(a) &(b)\\
\includegraphics[width=6.1cm]{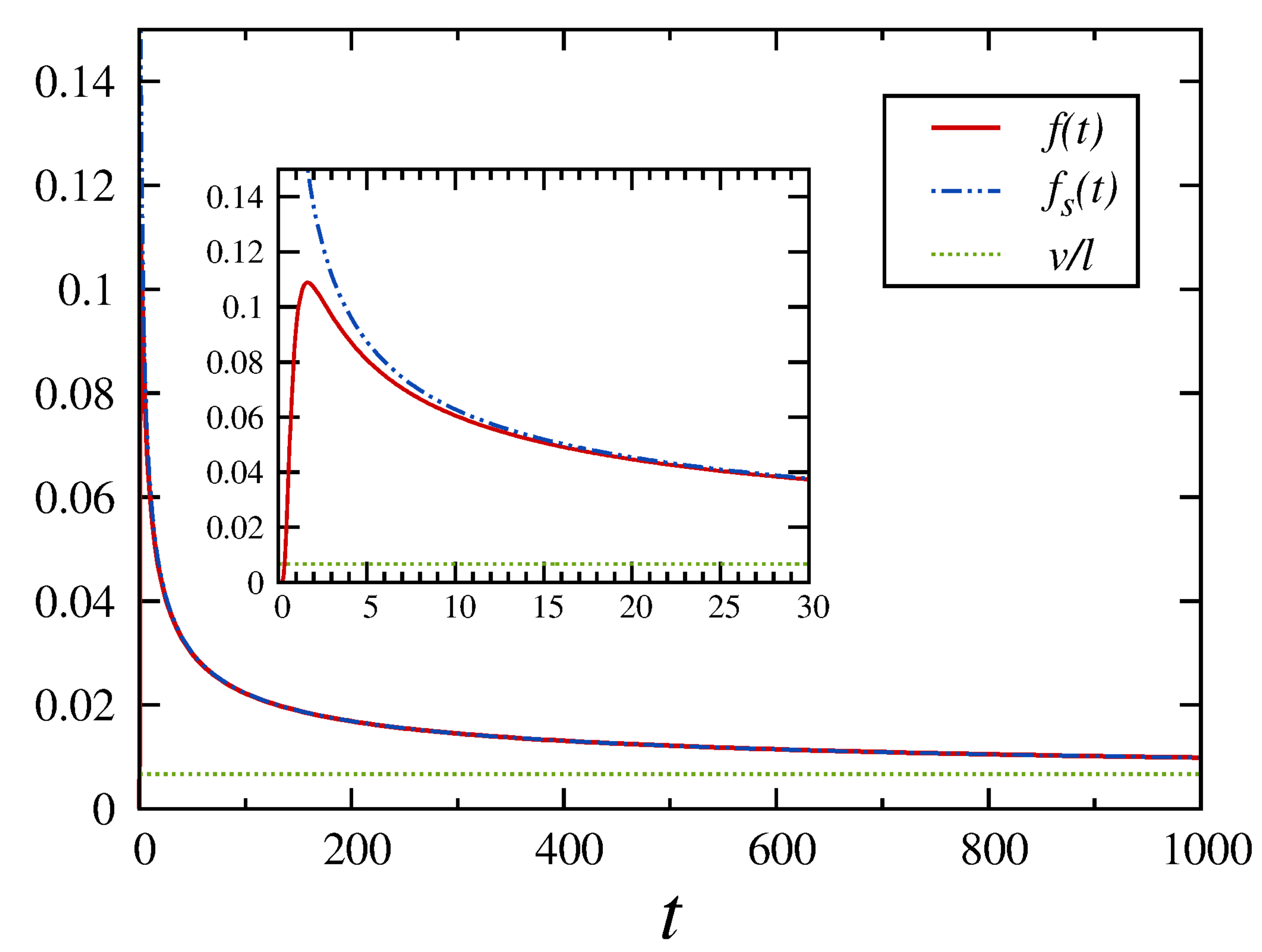} & \includegraphics[width=6.1cm]{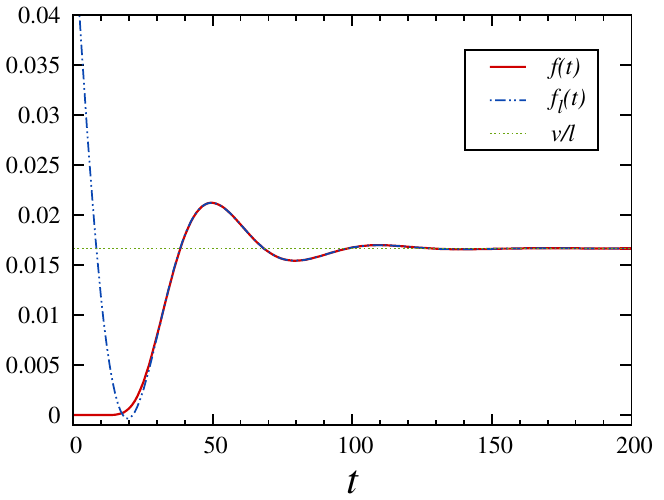} \\
\end{tabular}
\caption{ (a) The self consistent solution of $f$ and the approximation given by equation (\ref{smallvf}) for $P_e=0.03$, (b) The self consistent solution of $f$ and the approximation given by equation (\ref{largevf}) for $P_e=7.5$ } \label{fig2}
\end{center}
\end{figure}

In  figure \ref{fig2} we show the behaviour of the function $f$ for small and large P\'eclet number, together with the approximations given by equations (\ref{largevf}) and (\ref{smallvf}). The function $f(t)$  characterizes the relaxation to the stationary state, which in all cases  
is the  limiting value $v/l$. 

\section{Conclusion}

In this work we have considered a simple toy model where resetting to
the initial configuration is interaction driven. That is, the
resetting rate is not predetermined externally but is instead
determined self-consistently through the internal dynamics.  The toy
model comprises two Brownian particles biased to move towards each
other.  When the bias is strong and diffusion weak the resetting
becomes deterministic with fixed period however when the bias is weak
the resetting is dominated by diffusive effects.

Through a Green function technique, we have obtained an exact
expression for the non-equilibrium stationary state (\ref{sst})
(NESS).  This yields a non-trivial example of a NESS with a
probability current driven by internal resetting.  Furthermore, we have
obtained an exact expression for the full time-dependent distribution
(\ref{fsol}). This expression involves a function $f(t)$ (\ref{fntdef}) which
describes the effective resetting rate.  The function $f(t)$
characterizes the relaxation to the stationary state of the system at
large times: it is oscillatory in the high P\'eclet number regime
where diffusive effects are weak but monotonic in the small P\'eclet
number regime where diffusive effects are strong.  We have developed
simple distinct approximations to $f(t)$ in these two regimes.

It would be of interest to extend the toy model we have studied beyond a two particle system to a many particle system and to higher spatial dimension. It would also be of interest
to consider interaction-driven resetting in the context of search strategies of teams of searchers.

\ack 
RF thanks the kind hospitality at the Institute for Condensed Matter and Complex Systems, School of Physics and Astronomy, University of Edinburgh, where this work was done, and financial support from CNPq, Conselho Nacional de Desenvolvimento Cient\'ifico e Tecnol\'ogico-Brazil (233632/2014-0). 
MRE acknowledges partial support from EPSRC grant  EP/J007404/1.

\section*{References}

\providecommand{\newblock}{}

\end{document}